\documentclass[aps,prl,floatfix,twocolumn,superscriptaddress,showpacs,reprint]{revtex4-1}
\usepackage{graphicx}
\usepackage{mathrsfs}
\usepackage{amsmath}
\usepackage{subfigure}
\usepackage{bm}
\usepackage{verbatim}
\usepackage{color}
\usepackage{xcolor}
\usepackage[colorlinks=true, letterpaper=true, pdfstartview=FitV, linkcolor=blue, citecolor=blue, urlcolor=blue]{hyperref}
\newcommand{\mftc}{T_c^{\text{MF}}}
\newcommand{\cotc}{T_c^{\text{DW}}}

\newcommand{\vsl}{v_\text{SL}}
\usepackage{siunitx}

\begin{document}
\title{Spontaneous Layer-Pseudospin Domain Walls in Bilayer Graphene}
\author{Xiao Li}
\affiliation{Department of Physics, The University of Texas at Austin, Austin, Texas 78712, USA}
\author{Fan Zhang}\email{E-mail: zhf@sas.upenn.edu}
\affiliation{Department of Physics and Astronomy, University of Pennsylvania, Philadelphia, PA 19104, USA}
\author{Qian Niu}
\affiliation{Department of Physics, The University of Texas at Austin, Austin, Texas 78712, USA}
\affiliation{International Center for Quantum Materials, and Collaborative Innovation Center of Quantum Matter, School of Physics, Peking University, Beijing 100871, China}
\author{A. H. MacDonald}
\affiliation{Department of Physics, The University of Texas at Austin, Austin, Texas 78712, USA}
\begin{abstract}
Bilayer graphene is susceptible to a family of unusual broken symmetry states
with spin and valley dependent layer polarization.
We report on a microscopic study of the domain walls in these systems, 
demonstrating that they
have interesting microscopic structures related to order-induced topological characters.
We use our results to estimate Ginzburg-Landau model parameters and transition 
temperatures for the ordered states of bilayer graphene.
\end{abstract}
\maketitle

{\color{cyan}{\indent{\em Introduction.}}}---
Neutral bilayer graphene (BLG)~\cite{review,McCann} and its ABC-stacked multilayer
cousins~\cite{Min,Koshino,ABC-Zhang,SQH-Zhang},
are attractive platforms for unconventional two-dimensional electron systems physics 
because they have flat band contact near their Fermi levels, 
and because order induces large momentum-space Berry curvatures~\cite{SQH-Zhang} in their quasiparticle bands.  
Theoretical studies have identified a variety of potential 
broken symmetry states in neutral suspended BLG~\cite{SQH-Zhang,MF-Min,RG-Zhang1,RG-Zhang2,Vafek1,MF-Levitov,Falko1,Geim,MF-Jung,LAF-Zhang,MF-MacDonald,fRG,MC,
RG-Levitov,SQH-Levitov,Vafek2,Trushin1,Trushin2,Kharitonov,Gusynin,Vafek3,Vafek4,Falko2,Varma}.
The band eigenstates in bilayer graphene are equal weight coherent sums of components localized 
in each layer, and have an interlayer phase that is strongly wavevector dependent. 
When lattice-scale corrections to bilayer graphene's massive Dirac model~\cite{review,McCann} are neglected, 
the broken symmetry states predicted by mean-field theory have a 
charged quasiparticle energy gap~\cite{SQH-Zhang,MF-Min,RG-Zhang1,RG-Zhang2,MF-Levitov} and spontaneous layer polarization within each of the four spin-valley flavors.  
Recent experiments~\cite{Yacoby1,Yacoby2,Lau1,Lau2014,Schonenberger1,Lau2,Lau3,Wees,Schonenberger2}
appear to rule out a competing family of nematic states~\cite{Vafek1,Falko1,Geim},
which do not have a quasiparticle gap and break rotational symmetry~\cite{Cserti}.

The theoretical expectation~\cite{SQH-Zhang,MF-Min,MF-Jung,SQH-Levitov} is that long-range Coulomb interactions should favor the subset of broken symmetry states with no overall layer-polarization. 
Recent experiments~\cite{Schonenberger2}
utilize Zeeman response to an in-plane magnetic field~\cite{LAF-Zhang}
to identify the ground state as either a layer antiferromagnet~\cite{SQH-Zhang} in which 
opposite spins have opposite layer polarization, or a quantum spin Hall insulator~\cite{SQH-Zhang,SQH-Levitov}
in which layer polarization changes when either spin or valley is reversed.  (In
mean field theories the former state is favored by lattice scale exchange interactions~\cite{MF-Jung}.)  
In this Letter we present a microscopic theory of domain walls in which the sense of layer 
polarization of one flavor is reversed, focusing on the unusual properties associated
with the ordered states' topological characters. These domain walls 
proliferate thermally above an Ising phase transition temperature which we estimate and,
because they can be induced by spatial variations in the potential difference between layers,
are expected to be common in bilayer graphene samples.

{\color{cyan}{\it{Continuum model mean-field theory.}}}---
We first establish our notation by discussing uniform chiral symmetry breaking in BLG
in terms of the ordered state quasiparticle Hamiltonians~\cite{SQH-Zhang} suggested
by mean-field calculations and renormalization group 
analyses~\cite{MF-Min,RG-Zhang1,RG-Zhang2,MF-Levitov,LAF-Zhang,MF-Jung,MF-MacDonald,fRG}:
\begin{eqnarray}
\label{eq:generalmft}
{\mathcal H}^{HF}&=&\sum_{{\bm k}\alpha\beta s s'}c^{\dag}_{{\bm k}\alpha s}\big[h_0+h_{F}\big]
c_{{\bm k}\beta s'}\,,\nonumber\\
h_0 &=& -\epsilon_{\bm k} \big[\cos(2 \phi_{\bm k})\sigma_{x}^{\alpha\beta}+\sin(2 \phi_{\bm k})\sigma_{y}^{\alpha\beta}\big]\delta_{ss'}\,,\\
h_{F} &=& - \big[V_0+ V_{z}\sigma_{z}^{\alpha\alpha}\sigma_z^{\beta\beta} \big] \; \Delta_{\alpha s}^{\beta s'}\,\nonumber.
\end{eqnarray}
Here Greek letters label layer, $s$ and $s'$ label spin,
$\epsilon_{\bm k}$\,$=$\,${(\vsl \hbar k)^2}/\gamma_1$ is the band dispersion,
$\vsl$ is the single-layer Dirac-model velocity, $\gamma_1$ is the interlayer hopping energy,
$\cot\phi_{\bm k}$\,$=$\,$\tau_z k_{x}/k_{y}$ with $\tau_{z}=\pm 1$ denoting valleys $K$ and $K'$,
and $V_{0,z}$\,$=$\,$(V_{s} \pm V_{d})/2$ is  the sum and difference of the same ($s$) and 
different ($d$) layer interactions, which for convenience we assume to be short-ranged.  The order parameters
$\Delta_{\alpha s}^{\beta s'}$\,$=$\,$A^{-1}\sum_{{\bm k}}\langle c^{\dag}_{{\bm k}\beta s'}c_{{\bm k}\alpha s}\rangle_{f}$ must be determined self-consistently.  Note that in using short-range interactions
we are assuming that the screened Coulomb interaction range is short relative to 
the short-distance cut-off of the two-band continuum model, $\vsl \hbar / \gamma_1$, 
but much larger than the graphene lattice constant.  
The form used for the mean-field Hamiltonian in Eq.~\eqref{eq:generalmft}
has been simplified by noting that the mean-field ground state
has no net layer polarization, and that the mean-field interaction vertices 
are diagonal in layer~\cite{LAF-Zhang}.
This Hamiltonian generates a family of states differing only in the 
flavor dependence of the sign of interaction-generated mass terms proportional to 
$m_z \sigma_{z}^{\alpha,\beta}$.
In this Letter we concentrate on domain walls formed within a single flavor, reserving comments 
on the role of spin and valley degrees-of-freedom to the end of the article.

\begin{figure}[!]
\includegraphics[scale=0.50]{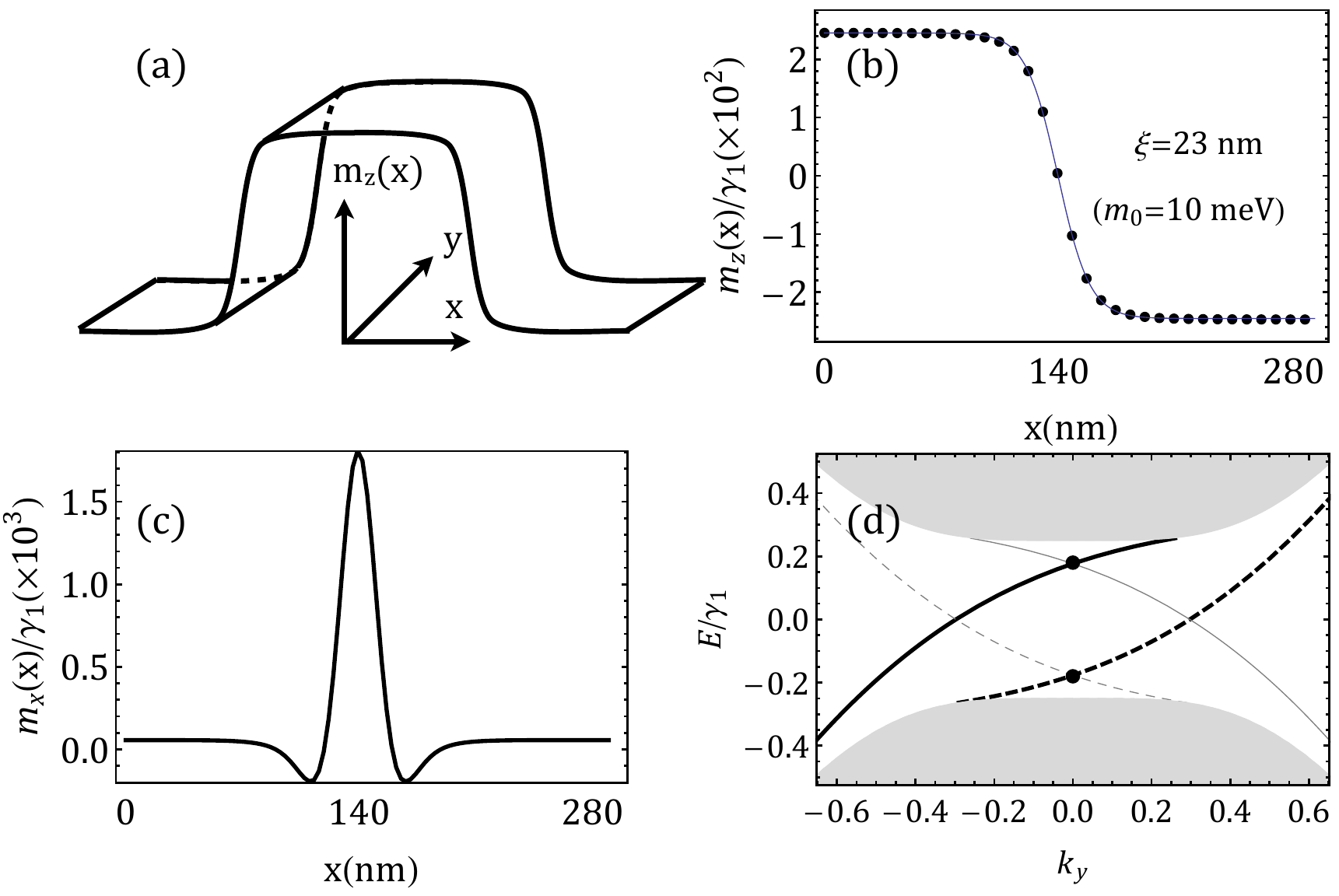}
\caption{\label{fig:one}(a) Schematic summary of our domain wall calculations.
These domain walls (kink and antikink) are oriented along the $y$ direction and the 
mass changes sign along the $x$ direction. 
(b)-(c) Typical mean-field solutions for $m_z(x)$ and $m_x(x)$ variation across a domain wall.
Note the different scales in (b) and (c).
(d) Energy spectrum of a model with sharp domain walls. 
The gray area is the bulk continuum.
Black and gray lines are used to distinguish chiral states 
localized at the domain walls which propagate in opposite directions while
solid and dashed lines are used to distinguish states with 
$\langle\sigma_x\rangle$\,$<$\,$(>)$\,$0$.
The two black dots identify the states with $E=\pm |m_0|/\sqrt{2}$
discussed in the text.}
\end{figure}

The gap equation can be solved to yield an implicit solution for $m_z$: 
\begin{align}
	1={\nu_0 V_s}\int_{0}^{\gamma_1}\dfrac{1}{2\varepsilon}[f(-\varepsilon-\mu)-f(\varepsilon-\mu)]d\varepsilon\,,
\end{align}
where $\nu_0$\,$=$\,$\gamma_1/(4\pi \hbar^2\vsl^2)$ is the band density-of-states per flavor,
$\gamma_1$ is the continuum model ultraviolet cutoff energy,
$\mu$ is the Fermi energy, $\varepsilon$\,$=$\,$\sqrt{\epsilon_{\bm k}^2+m_z^2}$,
and $f(\varepsilon)$\,$=$\,$(1+e^{\varepsilon/k_BT})^{-1}$ is the Fermi function.
For charge-neutral BLG and $m_z$\,$\ll$\,$\gamma_1$,
we find that the quasiparticle gap is $2m_z$\,$=$\,$4\gamma_1 \exp(-2/V_s\nu_0)$ at zero temperature,
and that $m_z$ vanishes at $T=\mftc$ where 
\begin{align}
\mftc=e^{\gamma}m_z/\pi k_B\,,
\end{align}
and $\gamma$ is Euler's constant.

{\color{cyan}{\indent{\em Microscopic theory of domain walls.}}}---
We now consider the microscopic electronic structure of the 
domain walls that separate regions with opposite layer-polarization signs.
These domain walls are quite different from those of an easy axis ferromagnet, for example,
because the layer-pseudospin dependent term in the band Hamiltonian is not a small correction
to an otherwise pseudospin independent Hamiltonian.   
In order to use periodic boundary 
conditions we must, as illustrated in Fig.~\ref{fig:one}(a), allow for
two adequately separated domain walls 
along the direction in which we allow the sign of mass to change.   
We use a plane-wave expansion method to solve the spatially inhomogeneous 
gap equations.  The interaction terms in the mean-field Hamiltonian 
are spatially local and can be 
parameterized in terms of position dependent masses $m_{i}(x)$ associated with the three Pauli matrices $\sigma_{i}$.  
For short-range interactions, their plane-wave matrix elements are
\begin{align}
m_{i}(k_{1}',k_{1})= \frac{V_s}{2A} \sum_{f\alpha\beta,{\bm q}} 
 \langle c^\dag_{k_{1}'\hat{x}+{\bm q},\alpha}\sigma_{i}^{\alpha\beta} c_{k_{1}\hat{x} +{\bm q},\beta}\rangle_f\,,
\end{align}
where $i=x,y,z$, and $f$ labels filled quasiparticle states. 
Note that the mass terms depend on $k_{1}'-k_{1}$ only.
The inverse Fourier transform of this function specifies $m_{i}(x)$.  

\begin{figure}[!]
\includegraphics[scale=0.61]{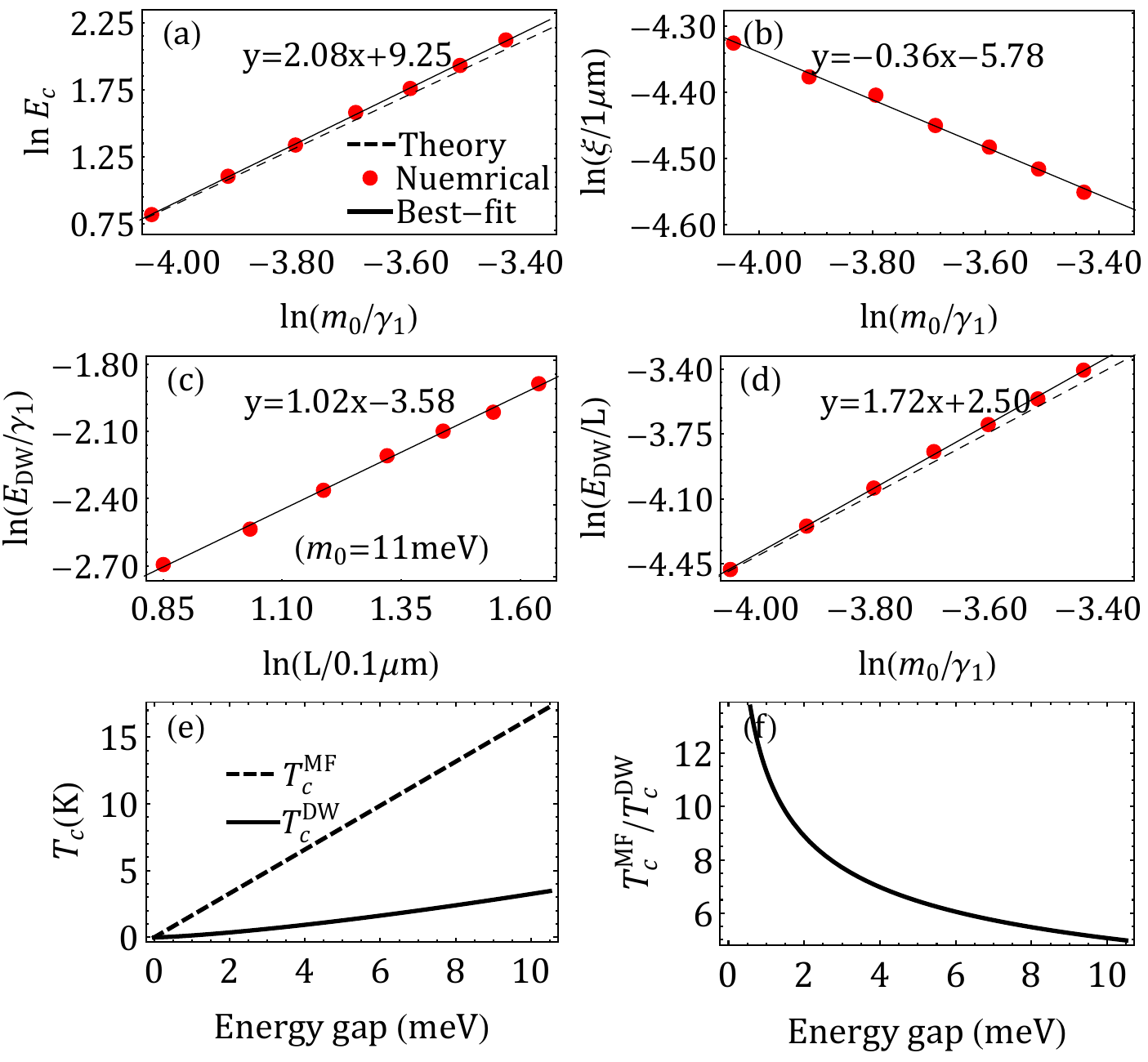}
\caption{\label{fig:two} Microscopic domain wall properties for square simulation cells with 
side $L$.  Calculations were performed as a function of $L$ and the interaction strength, and the results are plotted 
as a function of $L$ and the mass $m_0$ that interaction strength yields in the ground state.
The uniform system energy gap is $2 m_0$.   
In these figures, red dots are numerical data, while the thin solid lines are power law fits. 
(a) Condensation energy $E_{\text{cond}}$ of bilayer graphene (in units $\gamma_1/\mu\text{m}^2$) as a function of $m_0$. The dashed line is obtained from analytical results. 
(b) Domain wall width $\xi$ as a function of $m_0$. 
(c) Domain wall energy $E_{\text{DW}}$ as a function of $L$. 
(d) Domain wall surface tension $J\equiv E_{\text{DW}}/L$ (in units of $\gamma_1/0.1\mu\text{m}^2$) as a function of $m_0$. The dashed line is the GL theory prediction for the domain wall surface tension. 
(e) and (f) Comparison of the collective $\cotc$ and mean-field $\mftc$
critical temperature estimates discussed in the text.}
\end{figure}

These self-consistent gap equations are readily solved.
Results for finite square simulation cells of side $L$ are summarized in Fig.~\ref{fig:one} and Fig.~\ref{fig:two}.
A typical result for the domain wall $m_z$ profile, plotted 
in Fig.~\ref{fig:one}(b), can be accurately fit to the form  
$m_z(x)$\,$=$\,$m_0 \tanh[(x-x_0)/\sqrt{2}\xi]$, where $2m_0$ is the quasiparticle gap and  
$x_0$ is the position of the domain wall center.
As illustrated in Fig.~\ref{fig:two}(a) and (b), the energy cost of a domain wall $E_{\text{DW}}$ in our numerical calculation is accurately proportional to $L$,
indicating that finite-size effects are not playing a large role.
Fig~\ref{fig:two}(c) illustrates our finding 
that the domain wall energy per unit length (the two-dimensional  \emph{surface tension}) $J = E_{\text{DW}}/L$ 
and the domain wall width $\xi$ have power law dependences on the uniform system mass $m_0$:  
$ J \sim m_0 ^{\alpha}$ and $\xi \sim m_0^{\beta}$ with with $\alpha = 1.72$ and $\beta=-0.36$.
The surface tension increases and the domain wall width decreases with increasing $m_0$.  

{\color{cyan}{\it{Interlayer coherence response}}}---
The band states of bilayer graphene are coherent combinations~\cite{McCann} of top and bottom layer 
components with an interlayer phase $\phi$ that is twice the momentum orientation angle $\phi_{\bm k}$.
When represented by layer-pseudospins, valence band states are in the $x$-$y$ plane 
and have orientation angle $\phi = 2 \phi_{\bm k}$.  The $m_x$ and $m_y$ pseudospin magnetizations of both 
gapped and ungapped states therefore vanish after summing over momenta.   
As illustrated in Fig~\ref{fig:one}(c), our numerical calculations have revealed that a finite net in-plane pseudospin 
magnetization develops inside domain walls with a magnitude 
typically one order smaller than $m_0$.  The in-plane pseudospin magnetization is oriented across the domain wall,
{\em i.e.}, in the $x$-direction for the geometry we have chosen.  
Intriguingly, the sign of $m_x$ is the same for both kink and anti-kink domain walls.
This in-plane pseudospin magnetization cannot be understood in terms of gradient 
expansions based on uniform system quasiparticle linear response functions
since $\chi_{xz}({\bm q})=\chi_{yz}({\bm q}) =0$.

Near a domain wall, the sign of $m_z$ is reversed and the local Chern number changes by
two~\cite{SQH-Zhang,Martin,LSW}, giving rise to two chiral zero modes per valley
propagating along the domain wall, as illustrated in Fig.~\ref{fig:one}(d).
We attribute the finite $m_x$ value in the domain wall to the properties of the 
topological edge states it traps, as we now explain.
At any value of $k_y$ the mean-field Hamiltonian $\mathcal{H}$ in the presence of domain walls is invariant 
under simultaneous rotation by 180$^{\circ}$ around the pseudo spin $\hat{x}$ axis 
and mirror transformation $x \to -x$ through the domain wall: 
$\sigma_{x}\mathcal{H}\sigma_x$\,$=$\,$\mathcal{H}(-\partial_x,-x)$.
Here we assume that $x=0$ is chosen to lie at the mid-point of a 
single domain wall. It follows that for any $k_y$, the two components of the eigenstates $\psi(x)=[u(x), v(x)]^{\text{T}}$ satisfy $v(x)$\,$=$\,$\pm u(-x)$, and hence that the pseudospin operator $\sigma_x$ will have a non-zero expectation value near $x=0$.  
Similarly since $\sigma_y\mathcal{H}(k_y) \sigma_y$\,$=$\,$-\mathcal{H}(-k_y)$,
if $(u, v)^{\text{T}}$ is an eigenstate of $\mathcal{H}$ at $k_y$ with eigenvalue $E$, then
$(v, -u)^{\text{T}}$ is an eigenstate at $-k_y$ with eigenvalue $-E$.
It follows that the two chiral states with $E=0$ will appear at opposite values of 
$k_y$ and have opposite expectation values of $\langle\sigma_x\rangle$.
For example in the case of a sharp kink, {\em i.e.},  for $m_z(x)$\,$=$\,$m_0{\rm\,sgn}(x)$, the chiral states  
at $k_y$\,$=$\,$0$ have $E$\,$=$\,$\pm |m_0|/\sqrt{2}$ (lying in the gap) and $\langle\sigma_{x}\rangle$\,$=\mp1$. 
Although the edge states are not fully polarized in the general case, states within a given  
chiral state branch have non-zero values of $\langle\sigma_{x}\rangle$ with a common sign and 
the edge state occupations are generically different for any position of the chemical potential within the 
uniform state mass gap. 

Typical behavior is illustrated in Fig.~\ref{fig:one}(d).
The dashed and solid edge state branches have different signs of $\langle\sigma_{x}\rangle$  
and different occupations.   As a consequence, $m_x(x)$ exhibits a positive peak at each domain wall center.
This in-plane pseudospin magnetization is independent of the domain wall sign and valley index,
and thus survives summation over flavors for any gapped broken symmetry state that breaks chiral symmetry
within flavors~\cite{SQH-Zhang}.  We note that this nonlinear response also arises near electric field driven domain walls~\cite{Martin,LSW,Alden,Vaezi,Yao,Jung,Qiao,Peeters,Li} and layer stacking domain walls~\cite{LSW,Alden,Vaezi}.

{\color{cyan}\indent{\em Phenomenological theory of domain walls.}}---
The domain wall shape found in our numerical calculations is 
consistent~\cite{Lubensky,Tinkham} with the Ising-order 
Ginzburg-Landau-theory energy functional
\begin{align}
F= \int d^{2} \vec{r}  \left[\dfrac{c}{2}(\nabla m_z)^2+\mathcal{V}[m_z(x)] - E_c\right], \label{Eq:GL}
\end{align}
where $\mathcal{V}[m_z]$\,$=$\,$-r m_z(x)^2/2+um_z(x)^4$ with both $r$ and $u$ positive,
and $E_c$\,$=$\,$- r^2/16u$ is the condensation energy per unit area of the uniform $m_z$ ground state.
We include the constant $E_c$ in this expression so that the minimum value of $F$,
which occurs for constant masses $m_z^{\ast}$\,$=$\,$\pm m_0$\,$=$\,$\pm(r/4u)^{1/2}$, is zero. 
For a single domain wall configuration in which $m_z$\,$\to$\,$\pm m_0$ for $x$\,$\to$\,$\pm \infty$,
the functional~\eqref{Eq:GL} is minimized by $m_z(x)=\pm m_0\tanh[(x-x_0)/\sqrt{2}\xi]$
with $\xi$\,$=$\,$\sqrt{c/r}$.  The three independent parameters of the Ginzburg-Landau
model reproduce microscopic values for $m_0$, $\xi$, and $E_c$ when we set  
$c$\,$=$\,$4E_c\xi^2/m_0^2$, $r$\,$=$\,$4E_c/m_0^2$, and $u$\,$=$\,$E_c/m_0^4$.
In Fig.~\ref{fig:two}(a) we demonstrate that the GL theory expression for the domain wall
surface tension $J$\,$=$\,$8 \sqrt{2} \xi E_c/3$ agrees accurately with our microscopic calculations,
and that the power laws relating $\xi$ and $J$ to the microscopic gap satisfy 
$\alpha - \beta = 2$, also in agreement with the GL theory.

{\color{cyan}\indent{\em Ising critical temperature estimate}}---
We now utilize the above results to estimate the critical temperature $\cotc$ 
above which domain walls nucleated by thermal fluctuations proliferate and Ising long-range 
order within flavors is lost.  
For this purpose we follow a common physical argument~\cite{Lubensky} which compares the
energy cost associated with domain wall 
nucleation with the corresponding entropic free energy gain. 
The energy cost to form a domain wall with perimeter $P$ in the uniform state is $JP$,
whereas the entropy is $k_B\ln C_P$,
where $C_P$ is the number of distinct closed-loop 
non-intersecting $P/W$-step walks.  Here $W = 2\sqrt{2} \xi$~\cite{Lubensky}
is the minimum distance over which a domain wall can change direction.
Using $C_P=(1+\sqrt{2})^{P/W}$~\cite{Lubensky}, 
we find that for temperatures above $\cotc=WJ/(k_B \ln(1+\sqrt{2}))$,
the proliferation of domains separating regions with different layer polarization signs
is thermodynamically favored and long-range order is lost. 
Combining our numerical results for $\xi$ and $J$ yields
\begin{align}
\label{cotc}
	\frac{k_B \cotc}{m_0}  = \dfrac{0.64}{\ln(1+\sqrt{2})}(m_0/\gamma_1)^{\alpha+\beta-1}\,.
\end{align}
Since $\alpha+\beta-1 > 0$ and  $m_0 \ll \gamma_1$, $k_{B} \cotc \ll m_0$.  

We have so far ignored thermal fermionic  fluctuations which produce  
particle-hole excitations and would limit the critical temperature if the 
domain wall energy was very large.  Because the mean-field theory gap equation is identical to that 
of BCS theory, it implies a critical temperature limit that is proportional to $m_0$.     
As illustrated in Fig.~\ref{fig:two} (e) and (f), the ratio $\mftc/\cotc$ decreases with increasing $m_0$ 
in agreement with Eq.~\eqref{cotc}.   Noting that $\gamma_1 \sim \SI{400}{meV}$ and that 
experimental~\cite{Lau2} values of $m_0$ in bilayer graphene are always smaller
than $\SI{4}{meV}$, we conclude that the temperature to which spontaneous layer polarization order survives is  
limited in practice by domain wall nucleation.   

{\color{cyan}\indent{\em Discussion.}}---
It is instructive to compare spontaneously gapped bilayer graphene with BCS superconductors.
In both cases weak interaction instabilities lead to a linear dependence of $\mftc$ on the gap parameter $m_0$.
Collective properties differ qualitatively however, in the first place because of the difference
between the order parameter dimensions.  
In superconductors the collective excitations whose proliferation limits the critical temperature are vortices rather than domain walls.
Additionally the free fermion dispersion is linear near the Fermi energy in the 
superconductor case but {\it quadratic} in bilayer graphene.
As a result, the coherence length in superconductors is related to the gap $\Delta$ 
by $\xi \sim \hbar \vsl/\Delta$, and the collective limit on the temperature must therefore exceed the 
nucleation energy of a vortex, {\em i.e.}, $k_BT_c \sim E_c \xi^2\sim \varepsilon_F$, independent of and much
larger than the gap or the mean-field critical temperature estimate.     
A similar estimate of the collective limit on $T_c$ can be obtained by appealing to 
Kosterlitz-Thouless theory, which explains why critical temperatures of weakly disordered superconducting
thin films are accurately predicted by mean-field theory.  
In bilayer graphene on the other hand, the relationship between $\xi$ and the gap can be 
estimated using $m_0$\,$\sim$\,$(\hbar \vsl/\xi)^2/\gamma_1$.  This estimate yields 
$\beta = -0.5$, in rough agreement with the estimate $\beta$\,$=$\,$-0.36$ extracted from our 
numerical results.  It follows that for bilayer graphene, the collective fluctuation 
critical temperature estimate is comparable to mean-field-theory estimate, and becomes smaller
in systems with small gaps.  Unlike the case of superconductors, collective order parameter 
fluctuations play an important role in limiting the critical temperature in bilayer graphene. 

\begin{figure}[t!]
\includegraphics[scale=0.49]{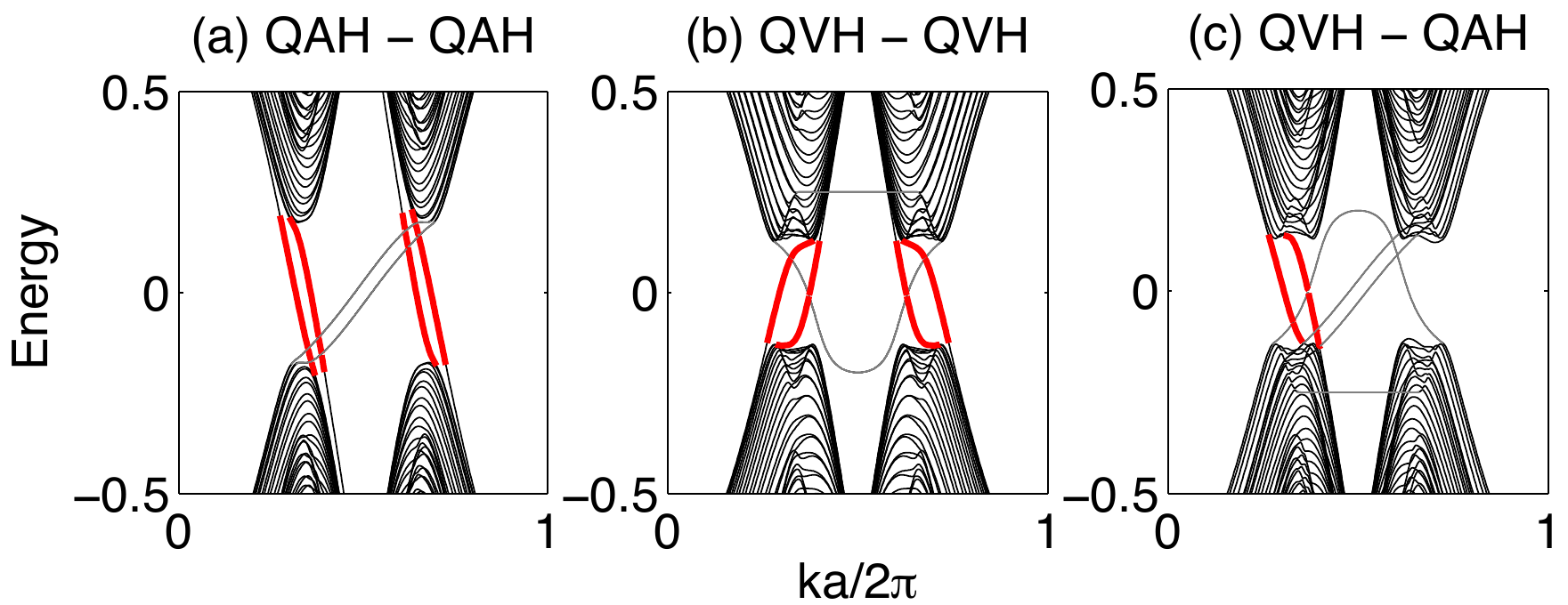}
\caption{\label{fig:three} Distinct domain wall zero-line patterns in gapped bilayer 
graphene samples which do not break spin-rotational symmetry.
The red lines denotes the zero modes localized at domain walls between (a) two QAH regions with opposite total Hall conductance,
(b) two QVH regions with opposite layer polarization, and (c) a QVH and a QAH region.
The gray lines represent the edge states on the outermost zigzag boundaries.  Note that they are doubly degenerate in (a) and (b). 
These figures were constructed using a tight-binding model of the gapped states.} 
\end{figure}

When the four spin-valley flavors and weak valley dependence of 
electron-electron interactions are taken into account,
the $2^4=16$ gapped broken symmetry states that are close in energy~\cite{MF-Jung} 
can be classified into $5$ distinct phases~\cite{SQH-Zhang}.
This in turn leads to $16$ distinct types of domain walls. 
In Fig.~\ref{fig:three} we illustrate only the cases in which spin rotational invariance is not broken.
In this case only the quantum valley Hall (QVH) state and quantum anomalous Hall (QAH) state are allowed, 
and support two intra-phase and one inter-phase domain wall.
Because the valley-projected Chern numbers 
are almost quantized~\cite{SQH-Zhang,Martin,LSW,Li} to $\pm 1$ in these 
states, all domain walls support chiral edge states.  
At a domain wall separating two QAH regions with opposite total Hall conductances, as 
illustrated in Fig.~\ref{fig:three}(a), the Chern number changes have the same sign for both 
valleys, yielding four modes with the same chirality.
At a domain wall separating two QVH regions with opposite layer polarization,
the Chern numbers change by $\pm 2$, with opposite signs for opposite valleys.
Thus two chiral zero modes (per spin) appear at valley K and two with opposite chirality at the valley K', as illustrated
in Fig.~\ref{fig:three}(b).  This type of domain wall can be easily realized using an external electric field~\cite{Martin,LSW,Alden,Vaezi,Yao,Jung,Qiao,Peeters,Li} 
or a stacking fault~\cite{LSW,Alden,Vaezi,Yao}.
Finally, at the domain wall between a QVH and a QAH regions the Chern number
is changed by two for one valley while it is unchanged for the other.
Thus the zero modes at the domain wall are chiral in one valley 
and absent in the other.  In all these cases, edge modes have a double spin-degeneracy.
States in which spin-rotational invariance is also broken can be similarly analyzed.
Each of the $16$ types of domain wall hosts a
Luttinger liquid~\cite{Affleck} with distinct properties.  
We emphasize that inter-valley scattering, ignored in the discussion above, 
should be extremely weak in the high quality samples required for the appearance 
of spontaneously gapped states, as the domain wall widths
we obtain are much larger than the graphene lattice constant.
Our work suggests that large-area bilayer graphene gapped states should 
exhibit interesting transport anomalies and their critical behaviors~\cite{LL}. 
Similar phenomena will occur in thicker ABC-stacked few-layer~\cite{SQH-Zhang}
graphene systems which have larger spontaneous gaps~\cite{Lau1,Lau2014} and
more robust domain walls.

{\color{cyan}\indent{\em Acknowledgements.}}---
X.L. acknowledges Z.~Qiao, H.~Chen and M.~Xie for helpful technical advice and is supported by NBRPC (No.~2012CB921300 and No.~2013CB921900) and NSFC (No.~91121004) during his visit at Peking University. F.Z. is indebted to C.~Kane and E.~Mele for helpful discussions and is supported by DARPA Grant No.~SPAWAR N66001-11-1-4110.  Q.N. is supported in part by DOE-DMSE (No.~DE-FG03-02ER45958) and the Welch Foundation (No.~F-1255). A.H.M. is supported by the Welch Foundation under Grant No.~TBF1473 and by the DOE Division of Materials Sciences and Engineering under grant No.~DE-FG03-02ER45958. 

\bibliographystyle{apsrev4-1}

\end{document}